\documentclass[12pt%
 preprint,
 amsmath,amssymb,
 aps,
]{revtex4-2}

\usepackage{xcolor}
\usepackage{graphicx}
\usepackage{dcolumn}
\usepackage{bm}
\usepackage{hyperref}
\usepackage{float}
\usepackage{ulem}
\usepackage{geometry}
\usepackage{comment}

\DeclareMathOperator{\sinc}{sinc}
\newcommand{\T}{\Tilde}
\newcommand{\x}{c_\psi \frac{\delta_x}{2}}
\newcommand{\sincx}{\sinc\left(\x\right)}
\newcommand{\sincxb}{\sinc\left(\overline{c}_\psi \frac{\delta_x}{2}\right)}

\newcommand{\xb}{\overline{x}}
\newcommand{\psib}{\overline{\psi}}
\newcommand{\hb}{\overline{h}}

\newcommand{\bs}{\boldsymbol}
\newcommand{\bo}{\boldsymbol}

\newcommand{\psip}{\dot{\psi}}

\newcommand{\xt}{\Tilde{x}}
\newcommand{\yt}{\Tilde{y}}
\newcommand{\zt}{\Tilde{z}}

\newcommand{\Vsub}{V_{\text{sub}}}
\newcommand{\Ssub}{S_{\text{sub}}}

\newcommand{\Basile}[1]{\textcolor{magenta}{#1}}

\begin{document}

\preprint{APS/123-QED}

\title{Flexible floaters align with the direction of wave propagation}

\author{Basile Dhote}
\email{basile.dhote@universite-paris-saclay.fr}

\author{Frederic Moisy}%
\author{Wietze Herreman}
\affiliation{Université Paris-Saclay, CNRS, Laboratoire FAST, 91405 Orsay, France \\}%

\date{\today}

\begin{abstract}

We investigate the slow, second order motion of thin flexible floating strips  drifting in  surface  gravity waves. We introduce a diffractionless model (Froude-Krylov approximation) that neglects viscosity, surface tension, and radiation effects. This model predicts a mean yaw moment that favors a longitudinal orientation of the strip, along the direction of wave propagation. The physical mechanism for this angular drift is analog to that of the standard linear Stokes drift: it originates from a slight imbalance between the stronger acceleration on the wave crests (that favors the longitudinal orientation) and the wave troughs (that favors the transverse orientation). Laboratory experiments with thin rectangular strips of polypropylene show a systematic rotation of the strips toward the longitudinal orientation, in good agreement with our model. We finally observe that the mean angular velocity toward the stable longitudinal orientation decreases as the strip length increases, an effect likely due to dissipation, which is not accounted for in our inviscid model.

\end{abstract}


\maketitle

\section{Introduction}

Modeling the motion of floaters in surface gravity waves is a classical problem in fluid-structure interaction with evident applications in ocean engineering \cite{newman2018marine,faltinsen1993sea,kim2008nonlinear} or microplastic transport \cite{suaria_2021,yang_2023,sutherland_fluid_2023,poulain_2024}. In low amplitude gravity waves, it is common to split the floater motion in a first-order harmonic response and a second order slow drift in both position and orientation angle (yaw angle). For solid floaters with fixed shapes, the theory is well established, particularly  in the inviscid, potential flow limit \cite{newman2018marine}.  The case of deformable floating bodies is more complex because they can change shape in response to waves, which is the primary focus of hydroelastic theory~\cite{bishop1979hydroelasticity,hirdaris2009hydroelasticity}. Such deformable structures include drifting nets \cite{mohapatra2024review}, agglomerated microplastic blobs \cite{yang_2023}, 
and may serve as models for oil spill patches \cite{huang_wave-induced_2013}. The study of very large and flexible floating structures has gained significant attention due to its relevance to ice floes \cite{meylan_response_1994} and emerging applications such as  offshore floating photovoltaic farms or floating airports \cite{zhang_2022}.

While most hydroelastic studies focus on the first-order harmonic response and the bending moment in the structure, some studies, such as Ref.~\cite{miao2019analysis}, also analyze the mean drift force and yaw moment acting on the structure. The slow second-order motion resulting from a second-order load on free, non-moored flexible structures has, however, been less frequently studied. Several experiments conducted with elongated flexible strips in waves have measured second-order drift in position \cite{kang_prediction_1996, wong_wave-induced_2003, huang_wave-induced_2013}, showing an increased drift velocity compared to the classical Stokes drift prediction for material points ~\cite{stokes1847theory,van_den_bremer_stokes_2018}. A slow drift in yaw angle was observed in the experiments of Wong {\it et al.}~\cite{wong_wave-induced_2003}, but without systematic measurements or modeling.

We have recently addressed the second order yaw motion of small elongated solid floaters drifting in propagating gravity waves~\cite{herreman2024}. Our experiments show that small solid floaters rotate toward a preferential orientation: short and heavy floaters align with the direction of wave incidence (longitudinal orientation), whereas light and long floater align parallel to the wave crests (transverse orientation).  This preferential orientation, either transverse or longitudinal, was first observed a century ago in the experiments of Suheyiro \cite{suyehiro1921yawing} using small boat models. Newman \cite{newman1967drift} calculated the mean yaw moment, but his model only predicts a transverse orientation for small floaters. In Ref.~\cite{herreman2024}, we calculated the mean yaw moment acting on rectangular parallelepiped floaters,  in the small floater limit, ignoring diffraction. We recovered Newman's contribution but also an extra yaw moment that favors the longitudinal position.  The changing preferential orientation is the result of these two opposing mean yaw moments. The physical origin of both mean yaw moments can be understood by considering the instantaneous moment. In crest positions, the instantaneous yaw moment always rotates the floater toward the longitudinal orientation, whereas in trough positions it pushes toward a transverse orientation. The mean yaw moment that favors the longitudinal state arises from the slightly higher amplitude in crest positions then in trough positions, as for the Stokes drift. The mean yaw moment that favors the transverse state is due to the variation of the submersion depth along the long axis of the floaters. Longer floaters have their tips more submersed in trough positions than in the crest positions and this significantly amplifies the instantaneous yaw moment in trough positions.  In the small floater limit of $kL_x \ll 1$, this inviscid and diffractionless theory suggests that the resulting preferred orientation of small solid floaters is governed by the non-dimensional parameter $ F = {kL_x^2}/{\hb}$, where $L_x$ is the floater length, $k$ is the wavenumber, and $\hb$ the equilibrium submersion depth. Solid floaters with $F < F_c$ (short or heavy) take longitudinal orientation, whereas for $F > F_c$ (long or light), the transverse orientation is preferred. The transition occurs at $F_c = 60$ in the case of parallelepiped floaters, in good agreement with our experiments in Ref. \cite{herreman2024}. 

This theory for solid floaters highlights how important the varying immersion depth is for the floater to rotate toward the transverse orientation. This naturally brings us to the question on how flexible floaters should orient in waves. Perfectly flexible floaters will deform and adapt to the shape of the interface, and are therefore not expected to exhibit variations in submersion depth. The mechanism responsible for the transverse orientation should be absent and so we expect a systematic longitudinal orientation for flexible floaters. The goal of this paper is to test this prediction.

In this paper, we introduce a model to describe the mean motion of a infinitely bendable thin floating strip drifting at the surface of inviscid water waves. Our model is inviscid and ignores diffraction  (Froude-Krylov approximation), as in Ref.~\cite{herreman2024}. It predicts a systematic longitudinal orientation of flexible floaters (along the direction of wave propagation). We also compute a small shape related negative correction to the classical Stokes drift.  We performed experiments with elongated rectangular, thin strips  of polypropylene or paper, and confirm the systematic longitudinal orientation. We measure the mean angular velocity toward the longitudinal orientation, and observe a decrease as the strip length increases. This effect, likely due to dissipation, is not accounted for in our inviscid model.

\section{Theory}

We introduce a model to describe the mean motion of a thin floating strip drifting at the surface of inviscid water waves. The strip is assumed to be infinitely bendable so that it adapts to the instantaneous  shape of the interface. Wave diffraction on  this adapting structure  will likely be negligible  because of the weak differential motion with the surrounding fluid flow. We can therefore use a  Froude-Krylov model, ignoring diffraction, to describe the action of the waves on the strip.

\subsection{Equations of motion}

We consider that the incoming wave is the classical  potential wave solution in infinitely deep water. The surface elevation and velocity potential are
\begin{eqnarray}
    \zeta & = & a \sin(kx - \omega t), \hspace{.5 cm} {\phi} = - \frac{a \omega}{k} e^{kz} \cos(kx - \omega t),
\end{eqnarray}
where $a$ is the wave amplitude, $k$ the wavenumber, and {$\omega = \sqrt{gk}$} the  frequency. The  fluid velocity is $\bs{u} = \nabla \phi$, and the pressure is $p = p_0 - \rho g z - \rho \partial_t\phi$, where $p_0$ is the atmospheric pressure, $g$ the gravitational acceleration and $\rho$ the fluid density.

\begin{figure}
    \centering
    \includegraphics[width = \linewidth]{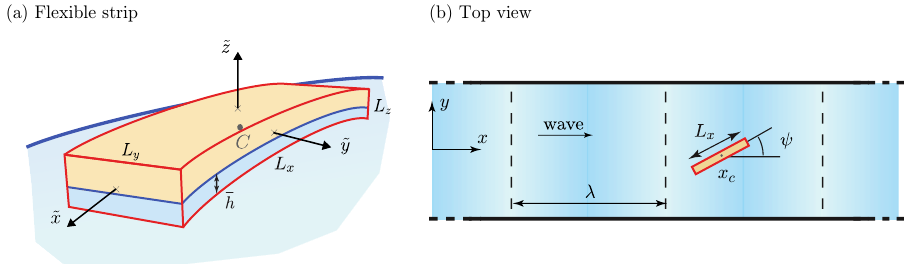}
    \caption{(a) Flexible strip with dimensions $L_x, L_y, L_z$, floating at the water surface with a constant immersion depth $\overline{h}$. (b) Top view of the strip drifting in a propagating wave. The yaw angle $\psi$ is the angle between the strip long axis and the direction of wave propagation.}
    \label{FIG1_ModelImages}
\end{figure}

The strip, sketched in Fig.~\ref{FIG1_ModelImages}(a), is a flexible rectangle of length, width and height ordered as $L_x \gg  L_y \gg L_z$, and density $\rho_{s} < \rho$. Although the final result pertains to a  thin  strip, we  need to  consider a finite thickness in order to express the buoyancy force and torque in terms of volume integrals. We assume that the strip has zero  bending  rigidity in the thin direction, but is inextensible in the other directions. We can therefore consider that the strip is everywhere immersed at the equilibrium submersion depth $\overline{h}$, as shown in Fig.~\ref{FIG1_ModelImages}(a). This equilibrium submersion depth is given by $\overline{h} =  \rho_{s}  L_z/\rho $ if only buoyancy is considered, but it can be slightly different because of capillary forces.

We parametrize the volume $V$ of the strip using curvilinear coordinates $(\T{x},\T{y},\T{z})$ along the long, middle and short axis, with $\T{x} \in [-L_x/2, L_x/2]$, $\T{y} \in [-L_y/2, L_y/2]$  and $\T{z} \in [- L_z/2,L_z/2]$. These coordinates are connected to the laboratory frame by 
\begin{eqnarray}
     x \approx x_c (t) + \xt \cos  \psi (t), \quad\quad y \approx y_c + \xt \sin \psi (t), \quad\quad z \approx  \zeta(x,t).
\label{tf}
\end{eqnarray}
We ignore the variation along $\T{y}$ and $\T{z}$ that are treated as very short directions ($L_y/L_x \ll 1$, $L_z/L_x \ll1$). The expressions above are valid up to corrections of order $O(\yt,\zt)$. The coordinates $x_c(t)$ and $y_c$ carry the horizontal position, and $\psi(t)$ denotes the yaw angle, i.e., the angle between the strip long axis and the direction of wave propagation, as shown in Fig.~\ref{FIG1_ModelImages}(b). The position $y_c$ does not change in time in this model. The submerged volume $\Vsub$ is parametrized using the same coordinates, except that $\T{z} \in [- L_z/2,- L_z/2 + \overline{h}]$. The velocity of a point $\bs{r}$ on the strip is
\begin{equation}
    \bo{v} \approx (\dot{x}_c - \psip \xt \sin \psi )\bo{e}_x  + \psip \xt \cos \psi \bo{e}_y + \partial_t \zeta \bs{e}_z.
\label{velocity_pt}
\end{equation}

Equations of motion for $x_c(t)$ and $\psi(t)$ are given by Newton's law ($x$-component) and the angular momentum theorem ($z$-component),

\begin{eqnarray}
    \frac{d}{dt} \bigl(M\dot{x}_c\bigl)& = & F_x,\quad \quad \text{with~~~}M\dot{x}_c = \int_{V} \rho_s v_x \, dV,\nonumber\\
    \frac{d}{dt}\bigl(I_{zz}\dot{\psi}\bigl)& =&  K_z, \quad\quad \text{with~~~} I_{zz}\dot{\psi} = \int_{V}\rho_s \, \bs{e}_z \cdot (\bs{r} - \bs{r}_c) \times \bs{v}  \, dV,
\label{Motion_eq}
\end{eqnarray}
with $M=\rho_s L_x L_y  L_z$ the mass and $I_{zz} \approx M L_x^2 /12$  the moment of inertia of the floater. We ignore capillary and viscous effects, and suppose that $F_x$ and $K_z$ is only due to the pressure of the incoming wave in the absence of floater. This hypothesis corresponds to the Froude-Krylov approximation, which neglects diffraction  and radiation.

Denoting $d\bs{S}$ the surface element on the floater toward the liquid, $\Ssub$  the submerged part of the floater surface and $\Vsub$ the submerged volume, we then have
\begin{subequations}
\begin{eqnarray}
    F_x  & = &  -\int_{\Ssub} (p - p_0) \bs{dS} \cdot \bs{e}_x  = - \int_{\Vsub}  \partial_x p \, dV  =  \int_{\Vsub} \rho \Big (  \partial_t u_x    + \underbrace{(\mathbf{u}\cdot\boldsymbol{\nabla})u_x}_{\sim\,\sin 2 (kx- \omega t)} \Big ) dV \label{Fx_vol_vanish}\\
     K_z & = &   -\int_{\Ssub} \left ( ( \bs{r} - \bs{r}_c ) \times (p - p_0) \bs{dS} \right ) \cdot \bs{e}_z   =   -\int_{\Vsub}  \rho (y - y_c)  \Big (  \partial_t u_x + \underbrace{(\mathbf{u}\cdot\boldsymbol{\nabla})u_x}_{\sim\,\sin 2 (kx- \omega t)} \Big ) \label{Kz_tocompute}   dV.
\end{eqnarray}
\label{Fx_surf_vol}
\end{subequations}
We have used the divergence theorem and Euler's law to rewrite $F_x$ and $K_z$ using volume integrals. Injecting the flow profiles, we can compute these integrals. The underbraced terms involving $(\mathbf{u}\cdot\boldsymbol{\nabla})u_x  \sim \sin 2 (kx- \omega t) $ are negligible in what follows. Of second order in wave magnitude, they do not affect the first-order motion of $x_c$ and $\psi$ and since they have a vanishing time-average, they do not influence the slow motion at second order.

\subsection{First- and second- order mean motion}

We scale space in units of $k^{-1}$, time in units $\omega^{-1}$, and mass in units $M$. {We denote $\epsilon = ka$ the wave slope, $\delta_{x,y,z} = kL_{x,y,z}$ the non-dimensional floater size, and $\overline{\delta}_{h} = k\hb$ the non-dimensional immersion depth. In non-dimensional units, the local fluid acceleration is $\partial_t u_x (x,z,t) = - \epsilon e^{z} \cos{(x - t)}$, and the surface elevation is $\zeta(x,t) = \epsilon \sin (x-t)$.  For thin elongated strips, we can approximate the volume integrals over $\Vsub$ as line integrals, yielding a non-dimensional force and moment}  
\begin{eqnarray}
F_x  \approx - \epsilon  \overline{\delta}_{h} \delta_y  \int_{-\delta_x/2}^{\delta_x/2}   e^{z} \, \cos{(x - t)} \, d\xt,\quad K_z  \approx  \epsilon \overline{\delta}_{h} \delta_y  \int_{-\delta_x/2}^{\delta_x/2}(y- y_c) \, e^{ z} \, \cos{(x - t)}  \, d\xt  .\label{Fx_lineintegral}
\end{eqnarray}
We replace $x,y,z$ with \eqref{tf} and notice that this force and moment can be rewritten in conservative form
\begin{equation}
{F}_x   \approx -  \frac{\partial V }{\partial x_c}, \quad \quad  {K}_z   \approx  - \frac{\partial V }{\partial \psi}  
\label{Forces_ConservativeForm}
\end{equation}
by introducing the potential
\begin{eqnarray}
V& =&  \overline{\delta}_{h} \delta_y \int_{-\delta_x/2}^{\delta_x/2}   e^{\epsilon \sin (x_c + c_\psi  \xt    -t  ) } \, d \xt,
\label{PotV_IntegralForm}
\end{eqnarray}
where we denote in short $c_\psi = \cos \psi$. We Taylor expand the exponential in the integrandum in powers of $\epsilon$ up to second order, and obtain
\begin{eqnarray}
V &\approx&    \overline{\delta}_{h} \delta_x \delta_y \left \{ 1 + \epsilon  \sinc  \left ( c_\psi  \frac{ \delta_x}{2} \right) \sin(x_c - t)    \right .
 \left .  + \frac{\epsilon^2}{4} \left [ 1  - \sinc  \left ( c_\psi  \delta_x \right)  \cos 2 (x_c - t)  \right ]   + O (\epsilon^3) \right \} .
\label{PotV_approximation}
\end{eqnarray}
The constant term $\overline{\delta}_{h} \delta_x \delta_y (1 - (\epsilon^2/ 4))$, does not create motion. The $O(\epsilon^2)$ time-dependent term $- (\epsilon^2/4)  \overline{\delta}_{h} \delta_x \delta_y \sinc  \left ( c_\psi  \delta_x \right)  \cos 2 (x_c - t)   $, only creates second-order harmonics but no second-order mean motion. We combine the equation of motion \eqref{Motion_eq} with the conservative forces \eqref{Forces_ConservativeForm} and their associated potential \eqref{PotV_approximation}, yielding
\begin{eqnarray}
         \Ddot{x}_c \approx - \frac{\partial}{\partial x_c} \left[ \epsilon   \sinc  \left ( \frac{ \delta_x}{2} c_\psi \right)  \sin(x_c - t)\right], \quad \qquad
         \Ddot{\psi} \approx - \frac{\partial}{\partial\psi} \left[ \epsilon {\frac{12}{\delta_x^2}}  \sinc  \left ( \frac{ \delta_x}{2} c_\psi \right)  \sin(x_c - t)\right].
    \label{eq_tot_xpsi}
\end{eqnarray}
We now perform a multi-scale analysis: we expand $x_c$ and $\psi$ in powers of $\epsilon$, and introduce multiple time scales on which they vary:
\begin{equation}
    x_c = \xb_c + \epsilon x'_c + O (\epsilon^2), \quad \psi = \psib + \epsilon\psi' + O (\epsilon^2)  ,\quad\dot{} \rightarrow \partial_t + \epsilon\partial_\tau + \epsilon^2\partial_{T}. 
    \label{MultiscaleAnalysis}
\end{equation}
The barred variables represent the floater's motion averaged over a wave period. The primed variables carry the harmonic response of the floater to the wave. Both primed and barred variables are $O(1)$, as are the time scales $t,\tau,T = O(1)$.  Injecting the expansion into the system and using Taylor series, we find at order $O(1)$ that $\partial_t\psib=0$ and $\partial_t\xb_c = 0$ : mean positions do not change on this time scale. At order $O(\epsilon)$, we find the differential equations for $x_c'$ and $\psi'$ that are readily integrated to 
\begin{eqnarray}
         x'_c = \sincxb \cos(\xb_c - t), \quad \quad  \psi'  =  {\frac{12}{\delta_x^2}}\frac{\partial}{\partial \overline{\psi}} \left[  \sinc \left (\overline{c}_\psi \frac{\delta_x}{2}\right ) \right] \sin(\xb_c - t),
    \label{eq_tot_xpsi_fast_integrated}
\end{eqnarray}
with $\overline{c}_\psi = \cos \overline{\psi}$.  Averaging the $O(\epsilon^2)$ problem over time, we find the equations for the slow variation of the mean variables:
\begin{eqnarray}
    \partial^2_{\tau\tau}\xb_c &=& 0 \hspace{1.5cm}\text{and}\label{slowxmotion}\\
    \partial^2_{\tau\tau}\psib &=&  -  \frac{\partial \overline{V} }{\partial \overline{\psi}},\quad \quad \text{with }
    \overline{V} = \frac{3}{\delta_x^2} 
    \left[ \underbrace{\sinc^2 \left(\overline{c}_\psi \frac{\delta_x}{2}\right) - 1}_{\text{due to }x'_c}
    \quad + \quad  \underbrace{{\frac{12}{\delta_x^2}}\left( \frac{\partial}{\partial \overline{\psi}} \left [ \sinc \left (\overline{c}_\psi \frac{\delta_x}{2}\right ) \right ] \right)^2}_{\text{due to }\psi'}\right].
    \label{slowpsimotion}
\end{eqnarray}

\begin{figure}
    \centering
    \includegraphics[width =\linewidth]{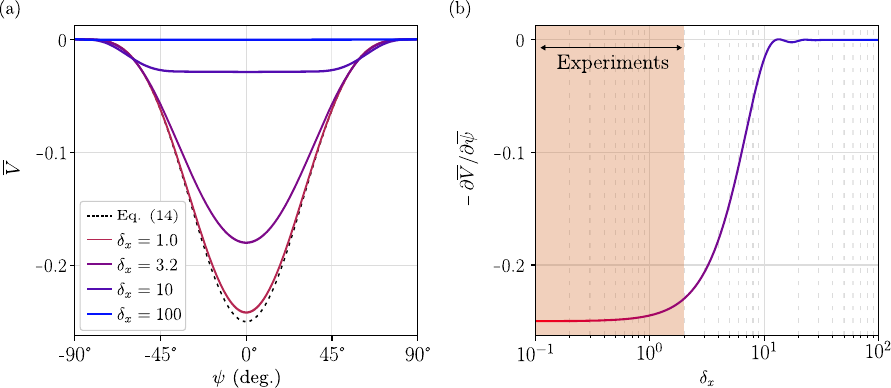}
    \caption{(a)  Effective potential $\overline{V}$ \eqref{slowpsimotion} governing the mean yaw angle for various strip lengths $\delta_x$. The black dotted line represents the small $\delta_x$ limit \eqref{shortlimit}. (b) Non-dimensional moment $- \partial\overline{V} / \partial \overline{\psi}$ for a mean yaw angle $\psib = 45^\circ$.}
    \label{FIG2_ThMoment}
\end{figure}

The equation on $\xb_c$ implies a constant drift velocity that is calculated in section C.
The slow acceleration of $\psib$ is therefore controlled by an effective potential $\overline{V}(\psib)$, which contains a contribution due to the fast back-and-forth displacement $x'_c$ and a contribution due to the fast yaw oscillations $\psi'$. The potential  $\overline{V}(\psib)$ is plotted in Fig.~\ref{FIG2_ThMoment}(a) for various strip lengths $\delta_x$. The minimum and maximum of $\overline{V}$ are always located at $\psib = 0^o$ and $\pm90^o$, demonstrating that the longitudinal orientation $\psib = 0^o$ is always stable and the transverse orientation is always unstable. The minimum at $\psib = 0^o$ is well pronounced for small $\delta_x$, suggesting a robust stability of the longitudinal orientation for small strips. However, as $\delta_x$ is increased, the  effective potential becomes gradually flatter, suggesting a weaker preference for the longitudinal orientation. This decreasing influence of waves on strips of large $\delta_x$ is better seen by computing the non-dimensional moment $- \partial \overline{V} / \partial \overline{\psi}$ for a fixed value of $\psib$, here taken equal to 45$^o$ [see Fig.~ \ref{FIG2_ThMoment}b]. It is systematically negative for short strips,  indicating a slow rotation toward the longitudinal orientation $\psib = 0^o$, for which we can offer a simple physical interpretation. During the passage of each wave crest, the strip experiences a negative torque, causing it to rotate longitudinally, in the direction of wave propagation. Conversely, at each wave trough, a positive torque rotates the strip transversely. Due to the asymmetry in dynamic pressure, which is slightly larger at the crests than at the troughs, a small net negative torque is generated, favoring the longitudinal orientation. For $\delta_x > O(10)$, i.e., for $L_x > 2 \lambda$, the moment becomes essentially zero with small oscillations, which can be seen as a canceling of the contributions of the large number of wave crests and troughs along the strip. 

If we integrate this inviscid model to find the motion of $\psib$, we will find slow  oscillations of $\psib$  around the stable fixed point $0^o$, with periods that depend on the initial conditions.
These periodic oscillations are not expected in a realistic experiment because there is viscous dissipation. Including dissipation in this model would yield damped oscillations converging toward $\psib = 0^o$.

At this point, no assumption has been made on the strip length $\delta_x$. The short-strip limit, relevant to the experiments presented in the next section, provides an interesting simplification of the problem. In this limit, the effective potential \eqref{slowpsimotion} becomes
\begin{equation} \label{shortlimit}
\overline{V} =  - \frac{1}{4} \overline{c}_\psi^4,
\end{equation} 
also plotted in Fig.~\ref{FIG2_ThMoment}(a) as the dotted line. In the short floater limit, there is a clear preference for the longitudinal orientation.  It is interesting to compare this effective potential for a flexible floater to the one obtained by  Herreman {\it et al.}~\cite{herreman2024} for a rigid floater,
\begin{equation} \label{shortlimit_rig}
\overline{V}_{\text{rigid}} =  -\frac{1}{4} \overline{c}_\psi^4 \left (  1 - \frac{\delta_x^2}{60 \overline{\delta}_{h}} \right ).
\end{equation}
In the rigid case, there is an extra term that can produce a change of sign of $\overline{V}_{\text{rigid}}$ when the non-dimensional parameter $F = \delta_x^2 /  \overline{\delta}_{h}$ (or $kL_x^2/\hb$) crosses the critical value 60. In other words, a long rigid floater with $F>60$ admits $\psib = 90^o$ (transverse orientation) as a stable equilibrium. We showed in Ref.~\cite{herreman2024} that this transverse equilibrium originates from the variation of the submersion depth along the floater length, which increases the torque in the trough positions, when the tips are more submersed. Because of the constant submersion depth for flexible floaters, this additional mechanism is absent, resulting in their systematic longitudinal orientation.

\subsection{Drift velocity}

In the previous section, we have found that $\overline{F}_x = 0 + O(\epsilon^3)$, resulting in $\ddot{\xb}_c = 0$. This suggests a constant drift velocity, which we compute as follows. We notice in the expression of the force $F_x$ \eqref{Fx_vol_vanish} that the time derivative can be taken outside the integral using Reynolds transport theorem.  Equation \eqref{Fx_vol_vanish} can therefore be written as 
\begin{eqnarray}
    F_x = \frac{d}{dt}\left(\int_{\Vsub} \rho u_x \, dV\right) + \int_{\Vsub}\rho \underbrace{\left((\boldsymbol{u} - \boldsymbol{v})\cdot\boldsymbol{\nabla}\right)u_x}_{\sim \, \sin(2(k\xb_c - \omega t))} \, dV. \label{Fx_tointegrate}
\end{eqnarray}
Here $\boldsymbol{v}$ is the velocity of the strip as defined in Eq.~\eqref{velocity_pt}. The second term of Eq.~\eqref{Fx_tointegrate} is of order $\epsilon^2$ and it can be shown, plugging in the first-order movement $(x'_c,\psi')$ of equations \eqref{eq_tot_xpsi_fast_integrated}, that it does not contribute to the mean motion at second order of the wave amplitude. The equation for the translation $m\ddot{x}_c = F_x$ is  then  integrated  once, yielding  
\begin{eqnarray}
    m\dot{x}_c = \int_{\Vsub} \rho u_x dV + cst.
    \label{cst_to_elim}
\end{eqnarray}
The integration constant is defined by the initial conditions, and can take arbitrary values in a perfect fluid.  However, by introducing a small amount of dissipation, the system would lose memory of the initial condition over a finite time, which justifies setting the integration constant to zero.
We nondimensionalize and calculate the integrals with the same techniques as before. This yields the equation
\begin{eqnarray}
    \dot{x}_c = \epsilon \sincx\sin(x_c - t) + \epsilon^2/2.
    \label{eq_psip}
\end{eqnarray}
We inject the expansion (\ref{MultiscaleAnalysis}) and Taylor expand the right hand side up to order $O(\epsilon^2)$. At order $O(\epsilon)$, we retrieve the same first-order movement $x'_c$ as found in Eq.~\eqref{eq_tot_xpsi_fast_integrated}. At second order, we inject the fast oscillating movement in translation ($x'_c$) and rotation ($\psi'$) in Eq.~\eqref{eq_psip} and we find the mean velocity
\begin{eqnarray}
    \partial_{T}\overline{x}_c & =& 1 + \frac{1}{2}
    \left[ \underbrace{\sinc^2 \left(\overline{c}_\psi \frac{\delta_x}{2}\right) - 1}_{\text{due to }x'_c}
    \quad + \quad  \underbrace{{\frac{12}{\delta_x^2}}\left( \frac{\partial}{\partial \overline{\psi}} \left [ \sinc \left (\overline{c}_\psi \frac{\delta_x}{2}\right ) \right ] \right)^2}_{\text{due to }\psi'}\right].
    \label{eq:xcdot}
\end{eqnarray}
We note that the time scale for the translational drift is $T$ in Eq.~\eqref{eq:xcdot} (i.e., second order in $\epsilon$), whereas that of the angular drift is $\tau$ in Eq.~\eqref{slowpsimotion} (i.e., first order in $\epsilon$). This  indicates that the slow angular dynamics is faster than the slow linear dynamics, suggesting that the floater reaches the longitudinal orientation in less than one wavelength.

The term in bracket in Eq.~(\ref{eq:xcdot}) is, with a factor $6/\delta_x^2$, identical to effective potential governing the angular drift (\ref{slowpsimotion}). It is always negative, and it contains the corrections to Stokes drift due to finite length and changes with angular position $\psib$. For very small floaters, in the limit $\delta_x \rightarrow 0$, the corrections vanish and we recover the classical Stokes drift velocity for a material point, $\partial_T\xb_c = 1$ (or $\epsilon^2 \omega / k$ in dimensional units). For infinitely long strips, we find $\partial_T\xb_c = 1/2$, exactly half of the material-point Stokes drift. This result can be understood as follows: for very long strips the fast oscillating first order motion \eqref{eq_tot_xpsi_fast_integrated} vanishes ($x'_c \rightarrow 0$ and $\psi' \rightarrow 0$). The strips do not move in the horizontal direction on short time scales. However they still deform vertically to adapt to the interface. Through this vertical motion, the vertical velocity gradients of the flow are seen by the strip and this results in a $1/2$ term that also appears in the derivation of the Stokes drift velocity of material points.

We note that the coupling between the slow evolution for the strip velocity and that of the yaw angle ($\partial_T\xb_c,\psib$) is unidirectional (equations (\ref{slowpsimotion}) and (\ref{eq:xcdot})). The drift velocity $\partial_T\xb_c$ depends on the mean yaw angle $\psib$, whereas the evolution of $\psib$ is independent of $\partial_T\xb_c$. To solve  this set of equations, we should therefore first integrate in time the mean yaw angle from Eq.~(\ref{slowpsimotion}), and then compute the resulting drift velocity $\partial_T\xb_c$ from Eq.~(\ref{eq:xcdot}). These coupled equations predict slow periodic oscillations of both $\psib$ and $\partial_T\xb_c$, at a frequency of order $\epsilon \omega$ that depends on the initial condition. The mean angle $\psib(t)$ slowly oscillates around the stable fixed point $0^o$ (longitudinal orientation), while the drift velocity $\partial_T\xb_c$ slowly oscillates around a mean value
smaller than the material-point Stokes drift $1$, with a lower bound at $(1+\sinc^2 (\delta_x/2))/2$.

Finally, in the small strip limit that applies to our experiments, the drift velocity (\ref{eq:xcdot}) simplifies to
\begin{eqnarray}
    \partial_T\overline{x}_c = \left(1 - \frac{1}{24} \delta_x^2 {\overline{c}_\psi^4} \right) + O(\delta_x^4).
\label{Vst_th_small}
\end{eqnarray}
This expression clearly shows a decrease of the drift velocity compared to the material-point Stokes drift prediction, which is more pronounced for a strip aligned along the direction of wave propagation.
However, this prediction of a reduced drift velocity is questionable in the presence of viscosity. Viscosity is known to {\it increase} the drift velocity of an extended flexible sheet~\cite{phillips1977dynamics, law_wave-induced_1999}, to an amount that may exceed the inviscid, shape-related reduction found here.

\section{Experiments}

\subsection{Experimental setup}

\begin{figure}
    \centering
    \includegraphics[width=\linewidth]{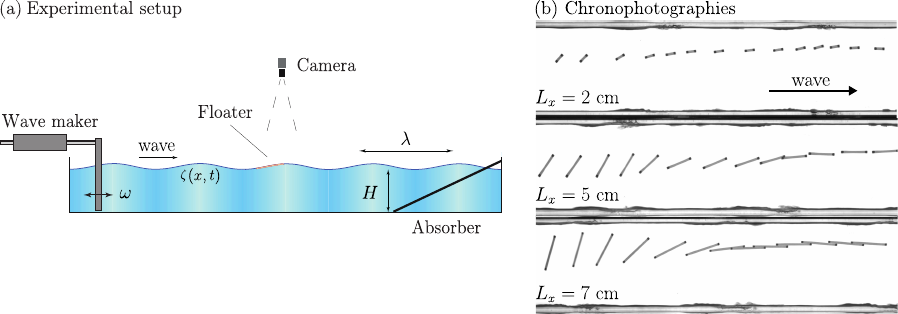}
    \caption{Experimental setup. (a) Side view. A $4$ m long, $18$ cm large wave tank is filled at height $H = 22$~cm. A wave maker oscillates at a frequency $\omega$, generating waves of velocity $c$ and wavelength $\lambda$. (b)~Chronophotographies of flexible strips of varying lengths $L_x$, taken at intervals of three wave periods, showing their systematic orientation along the direction of the wave propagation.}
    \label{FIG3_Chronophoto&Setup}
\end{figure}

We have performed a series of laboratory experiments to study the slow motion of flexible floaters drifting in surface waves. Experiments were conducted in a rectangular water channel sketched in Fig.~\ref{FIG3_Chronophoto&Setup}(a). The tank is $4$ m long, $18$ cm wide, and is filled with water at height $H = 22$ cm. The waves are generated by a piston driven by a linear motor, oscillating back and forth in a sinusoidal motion, and are attenuated at the end of the tank by a sloping plate of length 1.3~m and angle $12^\circ$.
The wave frequency $f = \omega/2\pi$ is varied between 1 and 2 Hz, corresponding to wavelengths in the range $\lambda \simeq 0.39 - 1.2$~m. In this range of frequencies, capillary effects can be neglected, and the waves obey the gravity-wave dispersion relation in finite depth, $\omega^2 = gk \tanh{kH}$. For each frequency, the wave amplitude $a$ and wavelength $\lambda$ are measured using a camera located on the side of the wave tank, with a precision of  $1$ mm and $1$ cm, respectively. Because of imperfect wave attenuation
at the sloping plate, the wave contains a small steady component, resulting in temporal oscillations of the amplitude of the order of 5\%.

Two sets of experiments were conducted. In the first, the frequency was fixed at 2 Hz (wavelength $\lambda = 0.39$~m), while the wave amplitude varied between 4 and 12 mm, yielding a wave slope $\epsilon \simeq 0.06 - 0.2$. In the second, the frequency was varied ($f= 1, 1.35$ and 2 Hz), with the amplitude adjusted to maintain a nearly constant slope, $\epsilon \simeq 0.14\pm 0.01$. This value is chosen to achieve a rapid drift and reorientation, while keeping the waves in the linear regime.

The floaters are rectangular strips of width $L_y = 5$ mm and lengths $L_x$ ranging from $1$ to $13$~cm (corresponding to $\delta_x = kL_x \simeq 0.1-2$), cut from polypropylene sheets of thickness $L_z\approx 53 \pm 1$ $\mu$m and surface density $\rho_\sigma =  48 \pm 1$ g/m$^2$.  At rest, such  thin strips lay flat at the water surface, due to the combined effects of capillary forces and rigidity. Capillary forces at the perimeter of the strip  induce a tension $\gamma \approx 0.07$ N/m, and the rigidity is characterized by the bending modulus
$$
D = \frac{EL_z^3}{12(1 - \nu^2)} \approx (1.5 \pm 0.2 )\, 10^{-6} \text{ N~m,}
$$
where $E$ is the Young's modulus, and $\nu$ the Poisson coefficient ($E\approx 0.09 \pm 0.01$ GPa and $\nu \approx 0.5$ for polypropylene). Gravity waves are unaffected by the tension $\gamma$ in the strip and its rigidity $D$ provided that~\cite{deike2013nonlinear}
$$
|\rho g \zeta| \gg |\gamma\partial^2_{xx}\zeta|, \, |D\partial^4_{xxxx}\zeta|.
$$
This requires wavelengths much larger than both the flexural-gravitary wavelength and the gravito-capillary wavelength, 
$$
\lambda \gg \lambda_{gD} = 2\pi \left( \frac{D}{\rho g} \right)^{1/4} \approx 2.2~\mbox{cm},\quad\text{and} \quad \lambda \gg \lambda_{gc} = 2\pi\sqrt{\frac{\gamma}{\rho g}} \approx 1.7~\mbox{cm}.
$$
With $\lambda$ in the range $0.4 - 1.2$~m in our experiments, these approximations are comfortably satisfied, indicating that the strips can be considered as infinitely bendable and that they perfectly adapt to the surface shape.

\subsection{Tracking and drift velocity}

Experiments to investigate the orientation dynamics and drift velocity of floating strips are conducted as follows. A strip is cautiously deposited on the water surface at rest (ensuring no air bubbles are trapped beneath it), at an initial yaw angle $\psi$ in the range $60-80^o$. The wave maker is switched on and the motion of the strip is recorded during approximately 1 minute using a camera located above the wave tank. We repeat the experiments several times for each strip length, and retain trajectories that remain approximately centered in the channel to discard possible interaction with the side walls. After each run, we wait approximately 10 minutes for the waves and currents to dissipate.

The chronophotographies in Fig.~\ref{FIG3_Chronophoto&Setup}(b), obtained by superimposing images at intervals of three wave periods, illustrate the motion of the strip for three different lengths, $L_x= 2, 5$ and 7~cm. We observe a reorientation of the strips in the direction of the wave propagation, in good agreement with the theoretical prediction. All experiments, carried out with varying strip lengths $L_x$ and wave frequencies $f$, systematically demonstrated this longitudinal orientation.

\begin{figure}
    \centering
    \includegraphics[width=0.9\linewidth]{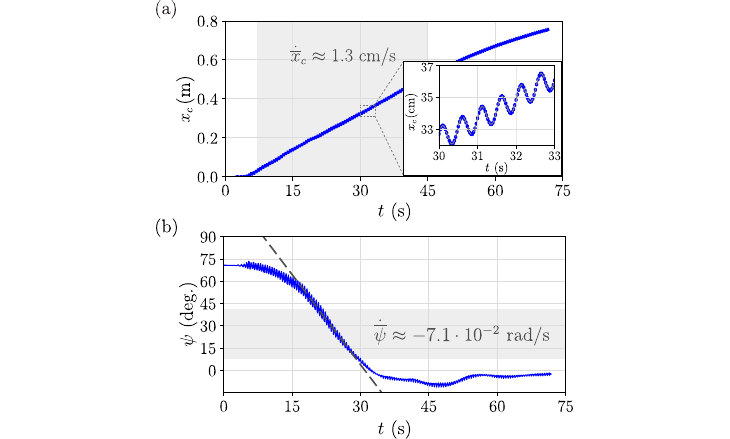}
    \caption{(a) Position and (b) yaw angle of a strip of length $L_x=8$ cm drifting in a wave of frequency $f = 2$ Hz, wavelength $\lambda = 0.39$ m, and wave slope $ak = 0.14$. The mean angular velocity is shown as a dashed line.}
    \label{FIG4_ExpData_Temporal}
\end{figure}

To further characterize the motion of the strip, we measure on each frame the position of its center of mass $x_c$ and its yaw angle $\psi$ using the TrackMate plug-in of ImageJ. This is achieved  by tracking the positions $(x_1,y_1), (x_2,y_2)$ of two black dots located at each end of the strip, from which we compute $x_c = (x_1+x_2)/2$ and $\psi = \tan^{-1} (y_2-y_1)/(x_2-x_1)$, with a resolution of $0.7$~mm for $x_c$ and $1^\circ$ for $\psi$. An example of time series of $x_c(t)$ and $\psi(t)$ is given in Fig.~\ref{FIG4_ExpData_Temporal} for a strip of length $L_x = 8$ cm. Both quantities clearly show rapid first-order oscillations at the wave frequency, superimposed to slow second-order drifts. The yaw angle shows a slow decrease, followed by small erratic motions of typically $\pm 10^\circ$ around the longitudinal direction $\psi = 0^\circ$.

In Fig.~\ref{FIG4_ExpData_Temporal}(a), we see that the strip rapidly acquires a nearly constant drift velocity $\dot{\xb}_c$ once the waves are established, approximately 5~s after the start of the wavemaker, but it significantly decelerates after $t \simeq 45$~s. This slower drift is likely due to the gradual development of a streaming flow, originating from the diffusion of vorticity produced in the oscillating boundary layers into the bulk~\cite{longuet1953mass,van_den_bremer_experimental_2019}. The resulting correction to the mean flow, which occurs typically after 45~s in our experiments, limits the duration of each run. In the following, the drift velocity is computed in this first time interval, where it is approximately constant. In the example shown in Fig.~\ref{FIG4_ExpData_Temporal}(a), we obtain  $\dot{\xb}_c \approx 1.3 \pm 0.2$~cm~s$^{-1}$, a value close to the expected Stokes drift of a point particle, $c \epsilon^2 \approx 1.5$~cm~s$^{-1}$, with $c = \omega/k$ the phase velocity (here $\dot{\xb}_c$ and $\dot{\psib}$, expressed in physical units, refer to the slow linear and angular drift velocities that superimpose to the rapid harmonic oscillations). This small discrepancy could be explained by the influence of the return flow, unavoidably present in a tank of finite depth: Mass conservation requires a zero flow rate through any vertical cross section, implying that a negative flow takes place to balance the positive Stokes drift~\cite{longuet1953mass,van_den_bremer_stokes_2018,Xiao_McAllister_Adcock_van_den_Bremer_2025}.  With the assumption of a homogeneous return flow, we expect a correction of $c\epsilon^2/(2kH \tanh{kH})$, resulting in a $10$ to $20\%$ reduction in the net Eulerian surface velocity. This is close to what we observe experimentally but may also be a coincidence: Eulerian return flows can become very complex and even time-dependent in closed wave-tanks.

\begin{figure}
    \centering
    \includegraphics[width=0.5\linewidth]{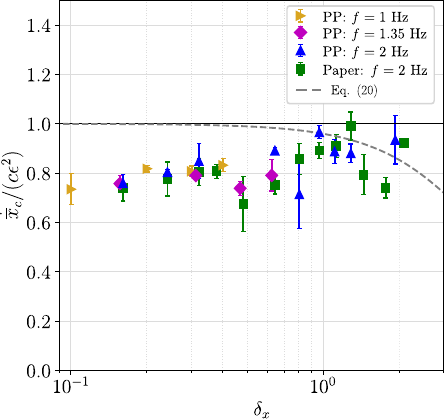}
    \caption{Normalized drift velocity $\dot{\xb}_c/V_{St}$ as a function of the non dimensional length $\delta_x = kL_x$ (PP: polypropylene). The full line at 1 is the expected material-point Stokes drift, and the dashed line is the predicted drift velocity for a floater in the longitudinal orientation [Eq.~\eqref{eq:xcdot} evaluated for $\psib=0^o$].}
    \label{FIG5_ExpData_MeanTranslation}
\end{figure}

We repeated the experiment for strips of various lengths and for various wave frequencies, and show in Fig.~\ref{FIG5_ExpData_MeanTranslation}(a) the drift velocity $\dot{\xb}_c/V_{St}$ normalized by the Stokes drift velocity as a function of the normalized length $\delta_x = k L_x$. Each point is calculated as the average over three to ten measurements, with error bars representing the standard deviations. Our measurements show a systematic deviation of $10-20\%$ below the material-point Stokes drift prediction $\dot{\xb}_c/V_{St} = 1$, consistent with the expected return flow contribution. In this figure we also show in dotted line the lower bound of the predicted drift velocity \eqref{eq:xcdot} for a perfectly longitudinal strip in infinite water depth, $\dot{\xb}_c/V_{St} = (1+\sinc^2 (\delta_x/2))/2$. In the range of strip lengths considered here, $\delta_x = 0.1 - 2$, this correction is less than 15\%, and  falls within the uncertainties due to the return flow and possibly to early streaming effects. We can conclude that the measured drift velocities, within the experimental precision, cannot be distinguished from the simple material-point Stokes drift prediction.


\subsection{Reorientation dynamics}

\begin{figure}
    \centering
    \includegraphics[width=\linewidth]{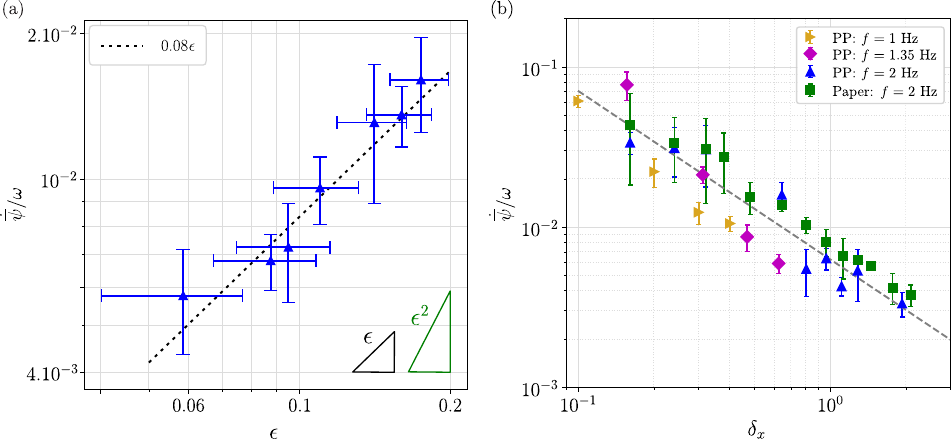}
    \caption{(a) Normalized angular velocity $\Dot{\overline{\psi}}/\omega$ as a function of the wave slope $\epsilon = ak$, for a wave frequency $f = 2$ Hz and a polypropylene floater of size $L_x=3$ cm. The dotted line shows best fit linear in $\epsilon$. (b) Normalized angular velocity $\Dot{\overline{\psi}}/\omega$ as a function the floater non-dimensional length $\delta_x$ (PP: polypropylene). The dashed line shows the power law \eqref{eq:fit}.}
    \label{FIG6_ExpData_MeanAngular}
\end{figure}

We now turn to the orientation dynamics of the strips. Our systematic observations of longitudinal orientation for all strip lengths are consistent with the theoretical prediction of a stable fixed point of the effective potential at $\psib = 0^o$. However, the theory predicts periodic oscillations around this stable fixed point, while experiments show a relaxation toward $\psib = 0^o$. This relaxation must originate from dissipation, not considered in the theory, likely  from the viscous boundary layer beneath the strips, or  to a lesser extent  from wave emission.

To characterize this relaxation dynamics, we measure the mean angular rate $\dot \psib$ by a linear fit of $\psi(t)$ in a time interval such that it decreases approximately linearly, as illustrated in Fig.~\ref{FIG4_ExpData_Temporal}(b). We observe that between $40^\circ$ and $10^\circ$ the decrease is linear and independent of the initial condition.  We note that this range occurs within the time interval during which the drift velocity is constant. The resulting (absolute) angular velocity $\Dot{\overline{\psi}}$ normalized by the wave frequency $\omega$ is plotted in Fig.~\ref{FIG6_ExpData_MeanAngular}(a) as a function of the wave slope $\epsilon = ak$ at a fixed frequency $f = 2$~Hz and for a floater of length $L_x = 3$~cm.  Here again, the vertical error bars represent the standard deviations from 3 to 10 experiments, and the horizontal bars represent the uncertainty in wave amplitude. Our data confirm that the angular drift scales as $\epsilon$, in good agreement with our theoretical prediction; this is in contrast with the drift in position, for which a $\epsilon^2$ scaling is predicted. 

In Fig.~\ref{FIG6_ExpData_MeanAngular}(b), the normalized angular velocity $\dot{\psib}/\omega$ is plotted for three wave frequencies as a function of the normalized length $\delta_x = k L_x$ for a fixed wave slope $ak \simeq 0.14$. We observe that $\dot{\psib}/\omega$ approximately decreases as $\delta_x^{-1}$ (a best fit with a power law $\delta_x^{-n}$ yields $n \simeq 1.1 \pm 0.2$). Together with the scaling in $\epsilon$ shown in Fig.~\ref{FIG6_ExpData_MeanAngular}(a), this suggests to write $\dot{\psib}/\omega$ in the empirical form
\begin{equation}
\frac{\Dot{\overline{\psi}}}{\omega} \simeq \alpha \, \epsilon\delta_x^{-1},
\label{eq:fit}
\end{equation}
where $\alpha \simeq (4.5 \pm 0.5)\, 10^{-2}$ is a fitting parameter. This scaling cannot be explained by our inviscid model, which predicts slow oscillations around the stable fixed point, characteristic of a second-order dynamics, whereas the experiments show a damping toward the fixed point, characteristic of a first-order dynamics. This suggests that the observed dependence \eqref{eq:fit} results from dissipation effects: longer floaters probably experience a stronger viscous drag, resulting in a slower angular dynamics. Confirming this hypothesis would require incorporating the effect of viscous dissipation into the model, which goes far beyond the scope of this study.

To check the robustness of the systematic longitudinal orientation, we have repeated the experiments with rectangular strips of paper, with the same width $L_y$ and range of lengths $L_x$ as before. The surface density of dry paper is $\rho_\sigma = 80$ g/m$^2$ but, contrary to polypropylene, paper absorbs water, so the relevant surface density is that of wet paper, $\rho_\sigma \approx 180 \pm 10$ g/m$^2$ (we estimate $\rho_\sigma$ by first soaking the paper strip in water and then weighing it, ensuring that no excess water droplets are present). We note that paper is heavier than water, so its flotation is ensured by capillary forces. The rigidity of wet paper  is comparable to that of polypropylene, $D \approx (1.1 \pm 0.5) \times 10^{-6}$ N~m, resulting in a similar flexural-gravitary crossover wavelength, $\lambda_{gD} \approx 2$~cm.  Experiments with these paper strips also show a systematic longitudinal orientation, with mean angular rates $\dot \psib / \omega$, plotted in Fig.~\ref{FIG6_ExpData_MeanAngular}(b),  in good agreement with that of polypropylene strips. This is consistent with our model, which predicts that the value of the immersion depth does not affect the angular dynamics of the strips, the only requirement being that they remain at the surface and deform with the waves.

Finally, we note that our results are also consistent with the qualitative observations of Wong \textit{et al.}~\cite{wong_wave-induced_2003} performed with  polyethylene sheets of elliptical shape. The authors also report a systematic trend for longitudinal orientation. From the images they provide in their paper for the parameters $f = 0.98$~Hz, $ak = 0.165$ and $\delta_x \approx 4.0$, we estimate a re-orientation rate $\dot{\psib}/\omega \approx 5.10^{-4}$. This value is consistent with our data, although twice smaller than our empirical fit (\ref{eq:fit}). This can be explained by the fact that floaters start from $\psib \approx 90^\circ$ in the study of Wong \textit{et al.}, and for these angles we expect a much smaller yaw moment.

\section{Conclusion}

In this paper, we introduced a Froude-Krylov model to describe the slow motion of a thin flexible strip drifting in surface gravity waves.  This model predicts a mean yaw moment that favors a longitudinal orientation of the strip, in the direction of wave propagation, and a small reduction of the drift velocity compared to the material-point Stokes drift. We performed laboratory experiments using strips of polypropylene of various lengths, that confirm the systematic longitudinal orientation, but cannot confirm the small reduction in drift velocity because of limited resolution. Although our experiments are performed for centimeter-scale strips, close to the capillary-gravity wavelength, capillary forces are not expected to play a significant role in the second-order forces and moments acting on the strip, suggesting that our results should hold for large-scale flexible structures.

The systematic longitudinal orientation of flexible floaters strongly differs from the behavior found for rigid floaters. While short and heavy rigid floaters also show a preferential longitudinal orientation, long and light rigid floaters orient transversely, parallel to the wave crests~\cite{herreman2024}. This transverse equilibrium arises from the variation of the submersion depth along the long axis of the floaters, which significantly increases the yaw moment in the trough positions. Since there is no variation in the submersion depth for perfectly flexible floaters, this additional mechanism is absent, resulting in their systematic longitudinal orientation. This difference between rigid and flexible floaters suggests a change in behavior governed by the bending rigidity of the floater. For floaters of intermediate rigidity, a situation relevant to large floating structures or ice floes~\cite{zhang_2022}, the hydroelastic response of the structure should result in a variable immersion depth along the floater, and we expect a behavior between the flexible and rigid limits, i.e., a longitudinal-transverse transition governed by the floater length, density, and rigidity.

Finally, our experiments suggest that the mean angular velocity of flexible floaters toward the stable longitudinal orientation decreases as $L_x^{-1}$, with $L_x$ the floater length. This effect is likely due to dissipation, and is not accounted for in our inviscid model, which predicts slow periodic oscillations around the longitudinal orientation. For small anisotropic particles that are neutrally buoyant, models based on Stokes flow theory are used to explain their preferential orientation~\cite{pujara2023wave, dibenedetto_orientation_2019}. Introducing dissipation in our model would require a detailed resolution of the second-order flow in the oscillating boundary layer beneath the floater. This approach was only considered for infinite floating sheets, yielding an increased drift velocity induced by the streaming in the oscillating boundary layer~\cite{phillips1977dynamics, law_wave-induced_1999}. Extending this approach to flexible floaters of finite size that are not aligned with the direction of wave propagation is necessary to provide a full description of their slow second-order angular dynamics, but would represent a considerable task.

\begin{acknowledgments}

We thank  A. Aubertin, L. Auffray, J. Amarni, E. Le Ster, and R. Pidoux for experimental help. This work was supported by the project “TransWaves” (Grant No. ANR-24-CE51-3840-01) of the French National Research Agency.

\end{acknowledgments}

\bibliography{apssamp}

\end{document}